%% file: main.tex
\documentclass[amsmath,twocolumn]{aastex702}

\newcommand{\teff}{\ensuremath{T_{\text{eff}}}}

\begin{document}

\title{Filling the Gap: Calibrating Gyrochronology at 1.3~Gyr with the Benchmark Cluster NGC-752}


\author[orcid=0000-0001-6037-2971]{Andrew W. Boyle}
\altaffiliation{NSF Graduate Research Fellow}
\affiliation{Department of Physics and Astronomy, The University of North Carolina at Chapel Hill, Chapel Hill, NC 27599, USA}
\email[show]{awboyle@unc.edu}

\author[orcid=0000-0003-3654-1602]{Andrew W. Mann}
\affiliation{Department of Physics and Astronomy, The University of North Carolina at Chapel Hill, Chapel Hill, NC 27599, USA}
\email{awmann@unc.edu}

\begin{abstract}

Empirical gyrochronology relies on co-eval stellar associations to map stellar rotation as a function of mass and age. The accuracy of its age predictions is therefore limited by the number and quality of benchmark rotation sequences. The well-characterized clusters NGC-6811 ($t\approx1$~Gyr) and NGC-6819 ($t\approx2.5$~Gyr) bracket an intermediate-age interval in which stellar spin-down remains poorly constrained due to lack of available rotation data. At $t\approx1.3$~Gyr, NGC-752 provides a critical benchmark within this gap, but previous measurements could not define the G- and K-dwarf rotation sequence needed for calibration. Here, we combine Gaia-based NGC-752 membership lists with literature rotation periods to map NGC-752's slow-rotator sequence for the first time. We identify a well-defined sequence intermediate between those of NGC-6811 and NGC-6819. Stars near $T_{\rm eff}\approx4500$~K rotate more slowly than their counterparts in NGC-6811, indicating that stars experiencing stalled spin-down at $\approx1$~Gyr have resumed appreciable spin-down by the age of NGC-752. \added{Adopting an age of 1.3~Gyr for NGC-752, we further show that existing empirical gyrochronology models overestimate the ages of many G and early-K dwarfs by 20--50\% because they lack intermediate-age calibration data, although uncertainty in the absolute age of NGC-752 may account for part of this offset.} NGC-752 therefore provides a necessary anchor for improving rotational age estimates. This calibration is especially timely because Gaia DR4, Roman, and PLATO are expected to yield rotation periods for tens of millions of stars, making the age coverage of benchmark sequences a principal limitation on large-scale gyrochronology.

\end{abstract}


\section{Introduction}\label{sec:intro}

Stellar rotation has been used to measure ages of cool, main-sequence stars, starting with the \citet{skumanichTimeScalesCA1972} relation showing that the rotational velocities of Sun-like stars decline with $t^{-1/2}$. This relation was formalized into the method of gyrochronology by \citet{Barnes2003}, which provides an empirical mapping between stellar rotation, \added{color}, and age. In the subsequent decades, gyrochronology has been recalibrated, primarily using a set of nearby benchmark clusters with well-determined rotation periods and independently determined ages. Today, \added{some of} the major benchmark clusters are $\alpha$ Per \citep[85\,Myr;][]{Boyle2023}, the Pleiades, \added{Pisces-Eridanus, and Blanco-1} \citep[120\,Myr;][]{rebullROTATIONPLEIADESK22016, curtisTESSRevealsThat2019, Gillen2020}, \added{NGC-3532 and Group-X \citep[300\,Myr;][]{Fritzewski2021, 2022A&A...657L...3M}, Praesepe and the Hyades \citep[670 and 700\,Myr, respectively;][]{Douglas2016,douglasK2RotationPeriods2019, rampalliThreeK2Campaigns2021}}, NGC-6811 \citep[1~Gyr;][]{meibomKEPLERCLUSTERSTUDY2011, curtisTemporaryEpochStalled2019}, NGC-6819 \citep[2.5~Gyr;][]{Meibom2015}, Ruprecht-147 \citep[2.7~Gyr;][]{curtisWhenStalledStars2020}, and M67 \citep[4\,Gyr;][]{Barnes2016}.

As the sample of stars with rotation periods has grown, it has become more clear that spin-down does not follow the single Skumanich-like law across all ages and masses. K~dwarfs in particular appear to reach the slowly-rotating sequence by 700\,Myr (Hyades and Praesepe) and then stall there for an extended period before resuming spin-down \citep{Agueros2018, curtisWhenStalledStars2020}. Separately, at older ages (\added{higher} Rossby numbers) stars undergo weakened magnetic braking \citep{vansadersWeakenedMagneticBraking2016}. M~dwarfs take much longer than their higher-mass counterparts to lose their initial rotation \citep[$>700$\,Myr and much longer for mid-to-late Ms;][]{curtisWhenStalledStars2020}, making it harder to map their spin-down law from existing benchmarks. These complications imply that gyrochronology cannot be mapped using a single scaling law \added{and make} it challenging to interpolate between cluster ages where even more undetected complications may hide. This underscores the need for denser observational coverage of clusters with ages between benchmarks and better mass sampling within known benchmark clusters.

Recent efforts have tried to move beyond analytic relations by mapping the full cluster-to-cluster variation and the scatter within a given cluster. \texttt{ChronoFlow} \citep{Van-Lane2025} uses a conditional normalizing flow trained on a large compilation of cluster rotators, while \texttt{gyro-interp} \citep{Bouma2023} achieves a similar effect by interpolating directly between empirical cluster sequences without assuming any specific functional form. Both approaches are limited by the number and quality of the benchmark clusters used to train them. 

NGC-6811 (1\,Gyr) and NGC-6819 (2.5\,Gyr) remain the two best-characterized clusters covering both K-dwarf stalling and slowed magnetic braking. Unfortunately, the 1.5\,Gyr gap between them is long compared to the timescales for observable changes in stalling and weakened magnetic braking \citep{Feiden2023}. The only nearby cluster with a well-established age in this interval is NGC-752 \citep[1.34\,Gyr;][]{Agueros2018}. However, NGC-752 has only 12 published rotation periods --- all for late-K and M stars, and as noted by \citet{curtisWhenStalledStars2020}, only eight of these are secure kinematic members. It remains a thinly sampled benchmark relative to its importance, and mapping NGC-752's rotation sequence for the F through early K-dwarfs could provide the missing calibration cluster necessary to fill this age gap.

TESS all-sky coverage has substantially expanded the census of rotation periods available in open clusters \citep[e.g.][]{Sha2024, Boyle2025b, Stafford2026}. We also now have longer-baseline surveys like the Zwicky Transient Factory \citep[ZTF;][]{Bellm2019}, which can recover periods for fainter and slower-rotators than TESS. The combination offers an opportunity to both densify the rotation sequences of existing benchmark clusters and to bring additional clusters into the gyrochronology calibration sample. 

In this paper, we present an expanded rotation sequence for NGC-752, combining new rotation periods from the TESS All-Sky Rotation Survey \citep[TARS;][]{Boyle2026}, deeper inspection of TESS data, and previously measured periods from \citet{Agueros2018}. Together, we bring the total number of secure rotators from eight to 25. 

\section{Membership List}\label{sec:mem}

\begin{figure}[!t]
    \centering
    \includegraphics[width=0.48\textwidth]{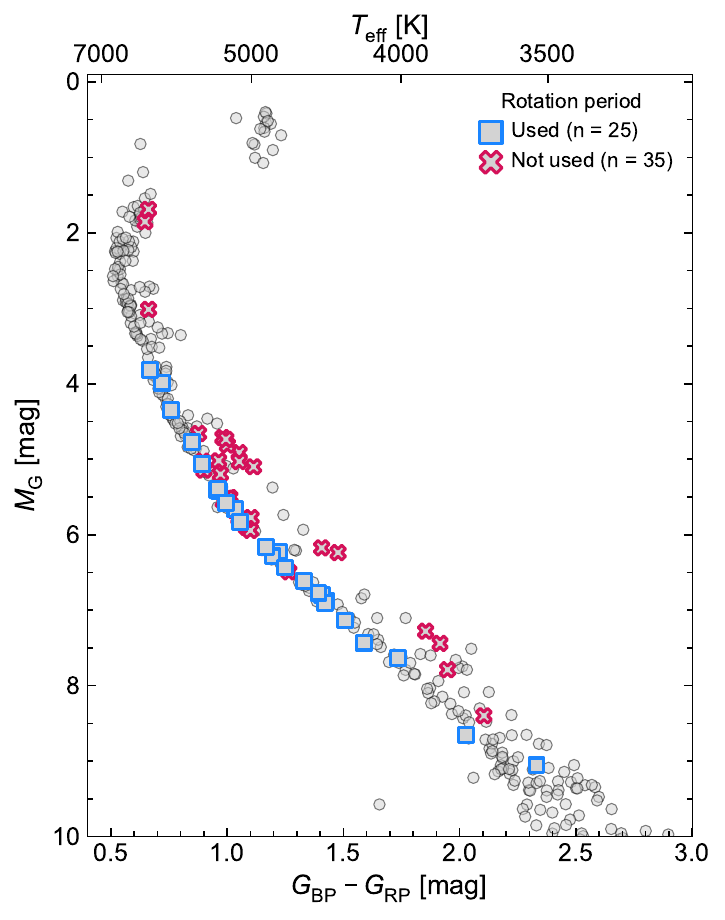}
    \caption{Gaia color-magnitude diagram of members of NGC-752 from \citet{Hunt2023}. Circles indicate catalog stars, red `X's are stars we excluded because of evidence of binarity or concerns about reliability (e.g., high RUWE, inconsistent periods between TESS sectors, high CMD position), and blue squares are stars with rotation periods used in this analysis. Approximate \teff{} is shown on the top for reference. }
    \label{fig:cmd}
\end{figure}

We opted to use the membership list from \citet{Hunt2023} and \citet{Hunt2024} for NGC-752. This list contains 442 probable member stars selected initially using an all-sky HDBSCAN clustering search on sources from Gaia DR3 \citep{GaiaCollaboration2021} combined with a neural-network based CMD classifier to assign ages, extinction, and distances. We show most of the Gaia color-magnitude diagram of NGC-752 using the \citet{Hunt2023} selection in Figure~\ref{fig:cmd}.

Several other catalogs contain additional candidate members, but we ultimately opted not to include these. \added{NGC-752 is known as Theia 1214 in \cite{Kounkel2020}, which contains} an additional 461 candidate members, but these are overwhelmingly offset in proper motion and XYZ position, exhibit much higher scatter in their CMD positions, and $<4\%$ of these stars have rotation periods in the TARS catalog (consistent with a random field draw). \citet{Agueros2018} has 258 candidate members, but 183 are within the \citet{Hunt2024} catalog, and nearly all the missing 75 have Gaia proper motions inconsistent with the cluster or are lacking astrometry in Gaia DR3. Lastly, \citet{Bhattacharya2021} has only six stars missing from \citet{Hunt2024}, none of which had usable rotation periods. 

\subsection{Extinction and Metallicity}
\added{
\citet{Agueros2018} reported an extinction toward NGC-752 of $A_V = 0.198^{+0.008}_{-0.009}$, whereas \citet{Hunt2023} reported $A_V = 0.07^{+0.04}_{-0.07}$. The default TARS catalog provides extinction estimates derived from the STILISM dust maps \citep{Vergely2022} for all stars with $T<16$ within 500~pc. We cross-matched the TARS catalog against the \citet{Hunt2023} membership list and calculated the median and 16th--84th percentile range of the resulting distribution. This procedure yielded $A_V = 0.11 \pm 0.01$.

NGC-752 has approximately solar metallicity. \citet{Agueros2018} measured $\mathrm{[Fe/H]} = 0.02 \pm 0.01$, while recent studies based on medium-resolution LAMOST spectroscopy found slightly subsolar values of $\mathrm{[Fe/H]} = -0.04 \pm 0.08$ \citep{2024A&A...692A.212Z} and $\mathrm{[Fe/H]} = -0.06 \pm 0.01$ \citep{2022A&A...668A...4F}.
}

\section{Rotation Periods}\label{sec:prot}

The default TARS catalog released with \cite{Boyle2026} contains over one million periods measured with Lomb-Scargle periodograms \citep{lombLeastsquaresFrequencyAnalysis1976, scargleStudiesAstronomicalTime1982}. These periods were vetted with two random forest classifiers. The first (the systematics classifier) separates quiet light curves and those dominated by TESS systematics from likely stellar variability, while the second (the harmonic classifier) identifies likely half-period harmonics in measured TESS periods. The combination of these two vetting steps allows periods as long as $\sim25$ days to be recovered from a single sector of TESS data.

\begin{figure*}
    \centering
    \includegraphics[width=1\linewidth]{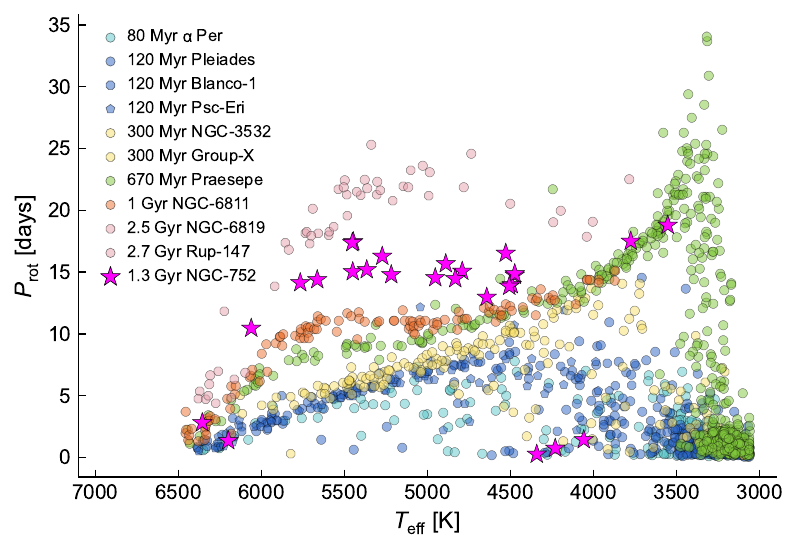}
    \caption{
    \textbf{The NGC-752 rotation-effective temperature sequence}.
    The NGC-752 rotation sequence with possible binary contaminants removed, plotted against other benchmark open cluster rotation sequences: $\alpha$ Per \citep{Boyle2023}, the Pleiades \citep{rebullROTATIONPLEIADESK22016}, Blanco-1 \citep{Gillen2020}, Pisces-Eridanus \citep{curtisTESSRevealsThat2019}, NGC-3532 \citep{Fritzewski2021}, Group-X \citep{2022A&A...657L...3M}, Praesepe \citep{rampalliThreeK2Campaigns2021}, NGC-6811 \citep{curtisTemporaryEpochStalled2019}, NGC-6819 \citep{Meibom2015}, and Ruprecht-147 \citep{curtisWhenStalledStars2020}. Each literature sequence is taken from \citet{Bouma2023} and only benchmark rotators are plotted. } 
    \label{fig:sequence}
\end{figure*}

NGC-752 lies 438~pc away \citep{Agueros2018}. At this distance, TESS photometry is inherently noisier than for nearby stars, causing stars with measurable rotation periods to score lower in the TARS systematics classifier. To counteract this effect, we regenerated the TARS catalog using the TARS catalog generation scripts available on Zenodo\footnote{\url{https://zenodo.org/records/19917941}} using a systematic classifier score threshold of 0.5. All other parameters were left the same as in \cite{Boyle2026}. We then cross-matched the list of 442 NGC-752 members from \cite{Hunt2023} with our TARS catalog, resulting in 75 matches. 

Eight of the twelve stars with published rotation periods in \citet{Agueros2018} for NGC-752 are also in the \cite{Hunt2023} NGC-752 membership list. We additionally adopted the \cite{Agueros2018} rotation periods for these eight stars.

\input{period_table}

Binary stars are a major source of contamination in stellar rotation studies as \added{close binarity can cause some stars to rotate faster than expected for their age}, resulting in the star appearing younger than it actually is \citep{Simonian2019}. We followed \cite{Bouma2023} and used the following criteria to remove likely binaries from our sample:

\begin{itemize}
    \item \textit{Gaia RUWE} --- We removed all stars with a RUWE $>$ 1.4.
    \item \textit{Gaia} \texttt{non\_single\_star} --- We removed all stars where Gaia's \texttt{non\_single\_star} flag did not equal zero.
    \item \textit{Overluminosity in the color-magnitude diagram} --- We manually selected and removed all stars that were elevated above the main sequence in $G$ vs. $G_{\rm BP} - G_{\rm RP}$, $G_{\rm BP} - G$, or $G - G_{\rm RP}$ (see Figure~\ref{fig:cmd}).
    \item \textit{Excess Gaia radial velocity uncertainty} --- Among stars with valid Gaia DR3 radial velocity measurements, we manually removed those with anomalously large uncertainties relative to stars of similar magnitude.
    \item For stars from the TARS catalog, we also applied the \texttt{flag\_possible\_binary == False} and \texttt{flag\_multiple\_periods == False} quality flags. The former flag is set to True when there are significant periodogram peaks at non-integer multiples of the adopted rotation period for that star, while the latter is set to True if any of the period measurements from the individual TESS sectors for a given star disagree with the adopted period for that star or its first order harmonics. 
\end{itemize}

We additionally removed nine stars from the TARS catalog that \added{passed the above criteria but} have temperatures greater than 6500~K since these stars are above the Kraft break so likely show a non-rotational type of variability. 

After applying binarity filters and quality cuts, we were left with 19 stars from TARS and seven stars from \cite{Agueros2018}, with one star in common between the two lists. This sample of 25 stars comprises our benchmark sample of rotators used to calibrate gyrochronology for NGC-752. These stars are shown on a color-magnitude diagram in Figure~\ref{fig:cmd}. All stars from the crossmatch between the \cite{Hunt2023} NGC-752 membership list and the TARS and \cite{Agueros2018} catalogs are reported in Table~\ref{tab:ngc752_periods} along with all quality flags we applied. An interested user could use the \texttt{flag\_benchmark\_sample == True} flag to isolate the stars that pass our quality cuts and comprise our benchmark sample of clean rotators.

\subsection{Period Validation}\label{subsec:prot_validation}

\begin{figure*}
    \centering
    \includegraphics[width=1\linewidth]{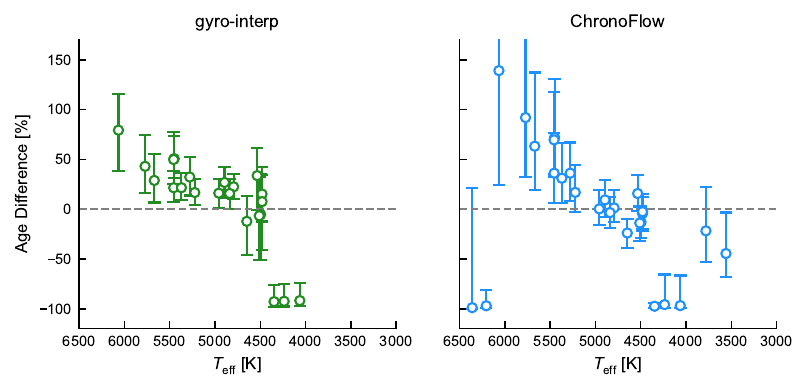}
    \caption{
    \textbf{Empirical gyrochronology models overestimate stellar ages around 1.3~Gyr due to lack of training data}. 
    \textit{Left:} The percent difference between \texttt{gyro-interp}-derived ages for our benchmark sample of NGC-752 rotators and the adopted cluster age. 
    \textit{Right:} The same as the left panel but with \texttt{ChronoFlow}-derived ages.
    An age difference of 50\% on this plot indicates the model overestimates the star's age by 50\%. Current gyrochronology models tend to overestimate stellar ages around our adopted age of $\simeq$1.3~Gyr due to the previous lack of a viable calibration cluster, create \teff-dependent systematics, and reduce overall precision. }
    \label{fig:age_difference}
\end{figure*}

Both of the source catalogs for our rotation periods describe their period validation in depth so we refer the reader to each of those papers for detailed period validation procedures. In addition to the period validation from each source catalog, we performed the following validation:

\begin{enumerate}
    \item \textit{External comparison} --- We selected stars from our regenerated TARS catalog with measured rotation periods between 13 and 18 days and effective temperatures between 4000 and 6000 K (roughly the same parameter space as NGC-752's slow sequence) and compared our adopted TARS periods to periods in the same parameter space from open cluster studies \citep{2023ApJS..268...30L}, K2 \citep{reinholdStellarRotationPeriods2020}, and ZTF \citep{Lu2022}. The periods in each of these studies are derived from larger telescopes with longer baselines than TESS so serve as an effective ``ground-truth'' against which we can compare our TESS periods. We applied the same cuts to the larger sample of TARS periods as we did in generating our sample of benchmark NGC-752 rotators (systematics classifier $>$ 0.5, harmonic classifier $>0.8$, \texttt{flag\_possible\_binary == False}, and \texttt{flag\_multiple\_periods == False}), with the exception of the photometric binary and excess radial velocity uncertainty binary cuts. This experiment showed that in this limited range, our TARS periods match the open cluster sample in $90\%$ of cases, the K2 sample in 82\%, and the ZTF sample in 94\%. In reality, the match rate of our benchmark NGC-752 rotator sample is likely higher due to our additional cleaning steps to remove binaries not included in this experiment.
    \item \textit{Internal comparison} --- There is one star in common between the TARS benchmark sample and the \cite{Agueros2018} benchmark sample: TIC~67461943. The TARS period for this star is $14.8 \pm 1.2$ days and the \cite{Agueros2018} period is $14.9 \pm 2.5$ days. Both periods agree within uncertainties, implying that TARS can effectively recover periods in this parameter space. 
    \item \textit{Visual examination} --- For each TESS sector used to calculate each TARS period, we manually examined the light curve, periodogram, and phase folded light curves at both the measured period and twice the measured period to ensure that our period estimates are visually identifiable.
\end{enumerate}


\section{The Gyrochronology sequence at 1.3~Gyr}

Figure~\ref{fig:sequence} places the rotation periods of our
benchmark NGC-752 members in the context of open-cluster
sequences spanning 80~Myr to 2.7~Gyr. \added{We plot each cluster's rotation sequence in terms of effective temperature, with each star's effective temperature estimated from de-reddened Gaia DR3 photometry following the procedure described in \cite{Boyle2025b}.} The primary gain in knowledge from this work over the \cite{Agueros2018} study is in defining NGC-752's slow rotator sequence by extending it to warmer temperatures. NGC-752's slow sequence appears to lie at $P_{\rm rot} \sim 15$ days and $4000 < T_{\rm eff} < 6000$~K. This flat morphology is consistent with the mass-dependent epoch of stalled spin-down identified by \citet{curtisTemporaryEpochStalled2019} in NGC-6811. In this picture, stars undergo an interval during which their surface rotation periods evolve only weakly after they arrive on the slow sequence, and the duration of this interval increases toward lower stellar masses. Hotter stars therefore resume spin-down earlier than cooler stars. By the age of NGC-752,
the hotter stars have had sufficient time to spin down toward the periods of cooler stars that remain near the stalled plateau, producing a nearly horizontal rotation--temperature
sequence.

The comparison with NGC-6811 makes this evolution particularly clear. Over approximately $4400\lesssim T_{\rm eff}\lesssim5000\,\mathrm{K}$, the NGC-752 slow sequence is displaced toward longer periods, implying that stars in this \teff{} range resumed appreciable spin-down during the few hundred megayears separating the two clusters. At the same temperatures, the NGC-752 sequence remains at shorter periods than the sequences of NGC-6819 and Ruprecht~147. NGC-752 therefore provides a direct empirical bridge between the $\approx1$ and $\approx2.5$~Gyr calibrators. The location of the stalled-spin-down transition at cooler temperatures is less well constrained; we do not have any data for stars on the slow sequence between $\sim3700$ and $4400$~K. The data are consistent with the stalled-spin-down boundary migrating toward cooler stars with age, but a larger sample of cool, apparently single members is needed to localize this boundary.

\added{
Three stars with $4000 < T_{\rm eff} < 4500$~K have rotation periods shorter than 2~days: TIC~67420277, TIC~67424741, and TIC~189573078. All three satisfy the single-star selection criteria described in Section~\ref{sec:prot}. However, stars at these temperatures should have converged onto the slow-rotator sequence by the age of NGC-752, making their rapid rotation unexpected.

We tested whether the measured periods could instead be caused by source contamination within the $21\arcsec$ TESS pixels. Neither the TIC v8.2 contamination ratio nor the TARS contamination quality flag, which identifies potential contaminants within $60\arcsec$ of each target, indicated significant contamination. We therefore find no evidence that the periods are spurious or originate from neighboring stars.

Having ruled out contamination as the likely explanation, we interpret the rapid rotation as evidence for unresolved binarity. Rapidly rotating stars at this age are predominantly tidally synchronized binaries \citep{Simonian2019}, but some companions could evade our selection criteria. In particular, Gaia RUWE becomes less sensitive to companions at projected separations below $\sim$100~mas or above $\sim$1000~mas, as well as to companions more than about five magnitudes fainter than the primary \citep{Wood2021}. We therefore conclude that these three stars are likely tidally spun up by undetected stellar companions.
}

\subsection{Empirical gyrochronology at 1.3~Gyr}

Although physics-based models of rotational evolution continue to improve \citep{2026arXiv260325792P}, empirical calibrations remain essential because the observed spin-down rate depends strongly on both stellar mass and age. Two publicly available empirical frameworks are \texttt{gyro-interp} \citep{Bouma2023} and \texttt{ChronoFlow} \citep{Van-Lane2025}. The \texttt{gyro-interp} calibration contains no benchmark cluster between NGC-6811 at $\approx1$~Gyr and NGC-6819 and Ruprecht~147 at $\approx2.5$--$2.7$~Gyr. \texttt{ChronoFlow} includes NGC-752, but its training sample contains only the eight rotation periods for stars with $T_{\rm eff} \lesssim 4500$~K reported by \citet{Agueros2018}, and therefore lacks a defined slow sequence. Thus, neither model has previously been constrained by a well-populated rotation sequence at the age of NGC-752.

Figure~\ref{fig:age_difference} shows the effect this lack of training data has on age predictions: both methods over-predict ages for the stars hotter than $\sim$5000~K in our benchmark NGC-752 rotator sample by more than $1\sigma$, implying that the derived age from these methods will often be $>400$~Myr away from the \added{adopted age}. Cooler stars remain stalled at this age, so predicted age medians lie close to the \added{adopted} value, albeit with relatively large uncertainties due to overlapping cluster sequences in the stalled regime. 

A similar concern is the loss of precision without the intermediate-age cluster. The gap in the rotation-period sequence from 1.0 to 1.3\,Gyr seen from $\simeq$4700\,K to 5700\,K indicates a continuing rapid spin-down over these 300\,Myr.  That implies a well-measured rotation period (small measurement uncertainties) could yield an age precision of $\lesssim200$\,Myr ($<15\%$). The empirically-calibrated tools discussed above yield age uncertainties of 20-50\% in this range due to the lack of calibration between 1\,Gyr and 2.7\,Gyr. 

\begin{figure}
    \centering
    \includegraphics[width=\linewidth]{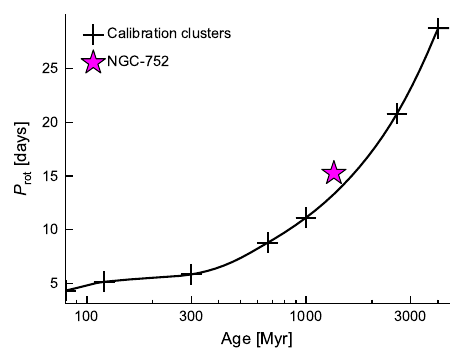}
    \caption{\textbf{Rotational evolution of a 5500~K star.}
    Adapted from Figure~4 of \citet{Bouma2023}. The black curve shows the
    rotation period evolution predicted by \texttt{gyro-interp} over the
    first ${\sim}4$~Gyr of a 5500~K star's life. Black plus signs mark the rotation periods at
    5500~K for the calibration clusters shown in
    Figure~\ref{fig:sequence}. The magenta star marks the representative
    location of NGC-752 at the adopted age of 1.3~Gyr. Its position above
    the model track indicates a longer rotation period than predicted,
    which could reflect an underestimated cluster age, rotational
    evolution not captured by the model, or both.
    }
    \label{fig:gyro_offset}
\end{figure}

\added{
Although we adopt an age of 1.3~Gyr for this study, recent age estimates for NGC-752 span a wide range. Isochrone-based analyses have yielded ages of approximately 1176~Myr \citep{2025ApJ...979...92W}, 1420~Myr \citep{Hunt2023}, and 1750~Myr \citep{2026A&A...706A..62M}. Using lithium measurements, \citet{2022ApJ...927..118B} found NGC-752 to be approximately twice as old as the Hyades, which itself has a lithium-depletion-boundary age of 695~Myr \citep{Galindo-Guil2022}. Finally, \citet{2023AJ....165....6S} derived an age of 1610~Myr from eclipsing binaries. This list is non-exhaustive but shows the current disagreement in the literature over NGC-752's age. 

Figure~\ref{fig:gyro_offset} shows the spin-down trajectory predicted by \texttt{gyro-interp} for a 5500~K star, together with the representative location of NGC-752 members at this temperature. At the adopted age of 1.3~Gyr, the NGC-752 sequence has a longer rotation period than predicted by the model. This offset could indicate that the adopted cluster age is underestimated, that the rotational evolution near this age is not fully captured by the model, or some combination of the two. Distinguishing among these possibilities is beyond the scope of this work and will require improved constraints on NGC-752's age.
}

\section{Summary and Conclusions}

We have used rotation periods from the TESS All-Sky Rotation Survey and \citet{Agueros2018} to map stellar rotation and calibrate gyrochronology at $\sim1.3$~Gyr with NGC-752. Our results are as follows:

\begin{enumerate}
    \item We have mapped a slow sequence in NGC-752, thereby providing a map of stellar rotation at a previously \added{undersampled} age range. 
    \item The comparison between NGC-752 and NGC-6811 indicates that stars with $4500\lesssim T_{\rm eff}\lesssim5000\,\mathrm{K}$, whose surface spin-down is stalled at $\approx1$~Gyr, have resumed appreciable spin-down by \added{the age of NGC-752}. This result supports a picture in which the transition out of stalled spin-down progresses toward cooler stars with increasing age.
    \item Existing empirical gyrochronology models show temperature-dependent biases when applied to NGC-752 due to the lack of available training data. In particular, both \texttt{gyro-interp} and \texttt{ChronoFlow} systematically overestimate the ages of many G and early-K dwarfs with $T_{\rm eff}\gtrsim5000\,\mathrm{K}$.
\end{enumerate}

We are entering an era where \added{large-scale} rotation period measurements will soon be available. Surveys like Gaia, Roman, and PLATO are expected to increase the number of stars with measured rotation periods into the tens of millions and enable gyrochronal age measurements for unprecedented stellar samples. However, the accuracy of these ages will remain limited by the available calibration data. Without benchmark clusters that densely sample stellar mass and age, systematic errors in the adopted spin-down relations will propagate into large-scale stellar age catalogs.

NGC-752 supplies one such benchmark at a previously undersampled age. Future progress will require similarly detailed rotation sequences for additional intermediate-age clusters, particularly among low-mass stars for which stalled spin-down produces large and asymmetric age uncertainties. Expanding this calibration set, while carefully accounting for membership, binarity, and period reliability, will be essential for translating upcoming rotation period catalogs into accurate stellar ages.

\begin{acknowledgments}

This material is based upon work supported by the National Science Foundation Graduate Research Fellowship Program under Grant No. DGE-2439854. Any opinions, findings, and conclusions or recommendations expressed in this material are those of the authors and do not necessarily reflect the views of the National Science Foundation. AWB thanks the LSST-DA Data Science Fellowship Program, which is funded by LSST-DA, the Brinson Foundation, the WoodNext Foundation, and the Research Corporation for Science Advancement Foundation; his participation in the program has benefited this work. 

AWM and AWB were funded by a grant from NASA's Astrophysics Data Analysis program (ADAP 80NSSC24K0619). AWM was further supported by a grant from the NSF CAREER program (AST-2143763).

This paper includes data collected by the TESS mission. Funding for the TESS mission is provided by NASA's Science Mission Directorate. The TESS data used in this paper can be found at MAST \citep{https://doi.org/10.17909/0cp4-2j79}.

This work has made use of data from the European Space Agency (ESA) mission
{\it Gaia} (\url{https://www.cosmos.esa.int/gaia}), processed by the {\it Gaia}
Data Processing and Analysis Consortium (DPAC,
\url{https://www.cosmos.esa.int/web/gaia/dpac/consortium}). Funding for the DPAC
has been provided by national institutions, in particular the institutions
participating in the {\it Gaia} Multilateral Agreement.

\end{acknowledgments}

\facilities{TESS \citep{Ricker2015}, Gaia \citep{2016A&A...595A...1G, 2023A&A...674A...1G}, Kepler \citep{Borucki2010}, K2 \citep{Howell2014}, ZTF \citep{Bellm2019}}

\software{astropy (\!\citealt{astropy:2013, astropy:2018, astropy:2022}),  
matplotlib \citep{hunter2007matplotlib},
pandas \citep{mckinney-proc-scipy-2010, reback2020pandas},
gyro-interp \citep{Bouma2023},
ChronoFlow \citep{Van-Lane2025}
          }

\bibliography{PAPER-Prot_Bayes, Mannbib}{}
\bibliographystyle{aasjournalv7.1}

\end{document}

%% file: period_table.tex
\begin{table*}
    \centering
    \caption{Rotation periods and quality flags for NGC 752 periods from TARS and \citet{Agueros2018}.}
    \begin{tabular}{ccc}
    \hline    \hline
    Parameter & Example Value & \textbf{Description} \\
    \hline
    \texttt{TICID} & 67424431 & TESS Input Catalog identifier \\
    \texttt{dr3\_source\_id} & 343133600227951616 & Gaia DR3 source identifier \\
    \texttt{period} & 15.722 & Adopted rotation period (days) \\
    \texttt{period\_unc} & 1.406 & Uncertainty on rotation period (days) \\
    \texttt{period\_provenance} & TARS & Source of the period measurement (TARS or \citealt{Agueros2018}) \\
    \texttt{teff} & 4894.9 & Calculated effective temperature (K) \\
    \texttt{ra} & 29.012114 & Gaia DR3 right ascension (deg) \\
    \texttt{dec} & 38.345883 & Gaia DR3 declination (deg) \\
    \texttt{pmra} & 9.5562 & Gaia DR3 proper motion in right ascension (mas/yr) \\
    \texttt{pmdec} & -11.5745 & Gaia DR3 proper motion in declination (mas/yr) \\
    \texttt{parallax} & 2.244 & Gaia DR3 parallax (mas) \\
    \texttt{phot\_g\_mean\_mag} & 14.531 & Gaia $G$-band apparent magnitude (mag) \\
    \texttt{phot\_bp\_mean\_mag} & 15.053 & Gaia $G_{\rm BP}$-band apparent magnitude (mag) \\
    \texttt{phot\_rp\_mean\_mag} & 13.858 & Gaia $G_{\rm RP}$-band apparent magnitude (mag) \\
    \texttt{phot\_g\_mean\_mag\_0} & 6.201 & Absolute $G$-band magnitude, corrected for extinction \\
    \texttt{phot\_bp\_mean\_mag\_0} & 6.701 & Absolute $G_{\rm BP}$ magnitude, corrected for extinction \\
    \texttt{phot\_rp\_mean\_mag\_0} & 5.547 & Absolute $G_{\rm RP}$ magnitude, corrected for extinction \\
    \texttt{BpmRp0} & 1.154 & Extinction-corrected $G_{\rm BP} - G_{\rm RP}$ color (mag) \\
    \texttt{extinction\_a0} & 0.104 & Monochromatic extinction at 550\,nm (mag) \\
    \texttt{ruwe} & 0.985 & Gaia DR3 Renormalized Unit Weight Error \\
    \texttt{non\_single\_star} & 0 & Gaia non-single star flag (0 = single, $>$0 = non-single) \\
    \texttt{flag\_multiple\_periods} & False & Inconsistent periods detected across sectors (TARS periods only) \\
    \texttt{flag\_possible\_binary} & False & Possible binary based on RUWE, Gaia flags, or light curve (TARS periods only) \\
    \texttt{is\_phot\_bin} & False & Photometric binary based on cluster CMD position \\
    \texttt{is\_rv\_bin} & False & Possible binary based on excess radial velocity uncertainty \\
    \texttt{is\_hunt\_member} & True &  Star is included in the \cite{Hunt2023} NGC-752 membership list \\
    \texttt{flag\_benchmark\_sample} & True & Passes all quality cuts; included in the benchmark rotator sample \\
    \hline
    \multicolumn{3}{c}{\footnotesize \textbf{Note.} This table is published in its entirety in machine-readable format. One entry is shown for guidance regarding form and content.}
    \end{tabular}
    \label{tab:ngc752_periods}
\end{table*}